\documentstyle[seceq,epsf]{ptptex}

\def\go{\mathrel{\raise.3ex\hbox{$>$}\mkern-14mu\lower0.6ex\hbox{$\sim$}}}
\def\lo{\mathrel{\raise.3ex\hbox{$<$}\mkern-14mu\lower0.6ex\hbox{$\sim$}}}

\title{
Darwin-Riemann Problems in Newtonian Gravity
}

\author{
Yoshiharu {\sc Eriguchi}$^{1,}$\footnote{E-mail address: 
eriguchi@valis.c.u-tokyo.ac.jp} 
and K\=oji {\sc Ury\=u}$^{2,}$\footnote{E-mail address: uryu@sissa.it}
}

\inst{
$^1$Department of Earth Science and Astronomy 
\\
Graduate School of Arts and Sciences, University of Tokyo 
\\
Komaba, Meguro, Tokyo 153-8902, Japan
\\
$^2$SISSA, Via Beirut 2/4, Trieste 34013, Italy}

\recdate{
}

\abst{
In this paper, we have reviewed the present status of the theory of 
equilibrium configurations of compact binary star systems in Newtonian 
gravity. Evolutionary processes of compact binary star systems due to 
gravitational wave emission can be divided into three stages according to 
the time scales and configurations.  The evolution is quasi-stationary 
until a merging process starts, since the time scale of the orbital 
change due to gravitational wave emission is longer than the 
orbital period.  In this stage, equilibrium sequences 
can be applied to evolution of compact binary star systems.  
Along the equilibrium sequences, there appear several critical states where 
some instability sets in or configuration changes drastically.  We have 
discussed relations among these critical points and have stressed the 
importance of the mass overflow as well as the dynamical instability of 
orbital motions.

Concerning the equilibrium sequences of binary star systems, we have summarized
classical results of incompressible ellipsoidal configurations. Recent results
of compressible binary star systems obtained by the ellipsoidal approximation
and by numerical computations have been shown and discussed. It is important
to note that numerical computational solutions to {\it exact equations}
show that compressibility may lead realistic neutron star binary systems
to mass overflows instead of dynamical disruptions for a wide
range of parameters.
}

\begin{document}

\maketitle

\section{Quasi-equilibrium stage of evolution of compact binary
star systems}
\subsection{Time scales and classification of evolutionary stages of 
compact binary star systems}
As is well known, gravitational waves from compact binary star
systems will carry away the angular momentum and energy from
the systems.  It implies that the binary star system will necessarily 
evolve to shrink its orbital radius and that eventually merging of two 
compact component stars will occur. There are three typical time scales 
to classify such an evolution: 1) the rotational period, $\tau_{\rm rot}$, 
2) the time scale of the orbital change due to gravitational wave emission, 
$\tau_{\rm GW}$, and 3) the light crossing time scale of the system, 
$\tau_{\rm dyn}$.

\begin{figure}
  \epsfysize = 5cm
  \centerline{\epsfbox{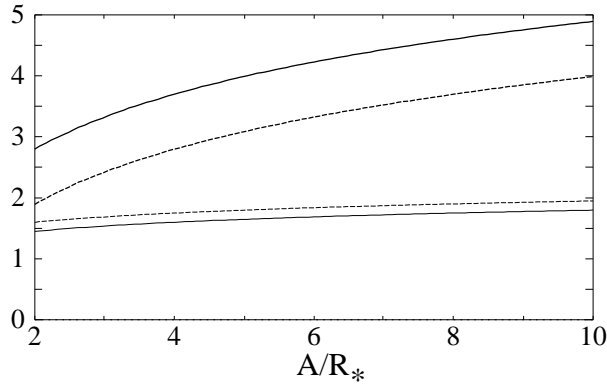}}
  \caption{Ratios of time scales $\log(\tau_{\rm GW}/\tau_{\rm dyn})$
  (thick curves) and $\log(\tau_{\rm rot}/\tau_{\rm dyn})$ (thin curves) 
  are plotted against the separation normalized by the stellar radius 
  for two values of $M/R_*$. $M/R_* =0.1$ (solid curves) and $0.2$ (dashed 
  curves).}
  \label{time scale figure}
\end{figure}

Let us consider a binary system consisting of two stars with masses $M_1$
and $M_2$. Two stars are in a circular orbit with a separation of 
$A \equiv 2d$. Here the separation of the binary system is defined by the 
distance between two centers of mass of two stars.  If stars are point masses 
with the same mass and the quadrupole formula for the gravitational radiation 
reaction is used, time scales mentioned above can be estimated as follows 
in Newtonian gravity:
\begin{eqnarray}
\tau_{\rm dyn} & \equiv    & d \ \ , \\
\tau_{\rm GW}  & \equiv & \left({1 \over A} {dA \over dt} \right)^{-1} 
          \approx {5 \over 8} \left({ M \over R_*}\right)^{-3}
                              \left({d \over R_*} \right)^3 \ \ , \\
\tau_{\rm rot} & \equiv & {2 \pi \over \Omega} 
          \approx  4 \pi \left({ M \over R_*}\right)^{-1/2}
                         \left({d \over R_*} \right)^{1/2} \ \ ,
\end{eqnarray}
where $R_*$ and $\Omega$ are the radius of the star and the angular velocity,
respectively, and geometrized units $c=G=1$ are used. If we specify the value 
of the compactness $M/R_*$, we can evaluate
these time scales as shown in Table~\ref{time scale table} and 
Fig.~\ref{time scale figure}. In Table~\ref{time scale table} ratios
of time scales, the rotational velocity $v_{\rm rot}$ and the number of 
rotations during the time scale $\tau_{\rm GW}$, $N$, are shown for two 
cases of the compactness $M/R_*$. In Fig.~\ref{time scale 
figure}, the ratios of time scales are plotted against the separation 
normalized by the stellar radius.
\begin{table}
\caption{Ratios of time scales, the rotational velocity $v_{\rm rot}$
and the number of rotations, $N$, during $\tau_{\rm GW}$ at the distance
of several stellar radii for different compactness parameters.}
\label{time scale table}
 \begin{center}
  \begin{tabular}{cccccr}
   \hline
   \hline
   $ M /R_*$ & $d / R_*$ & $\tau_{\rm GW} / \tau_{\rm dyn}$ 
    & $\tau_{\rm rot} / \tau_{\rm dyn}$ & $v_{\rm rot}/c$ & $N$ \\ 
   \hline
   0.1  & 3.0 & 1.7E4 & 6.9E1 & 0.09 & 250 \\
        & 2.0 & 5.0E3 & 5.6E1 & 0.11 &  90 \\
        & 1.0 & 6.3E2 & 4.0E1 & 0.16 &  16 \\
   \hline
   0.2  & 3.0 & 2.1E3 & 4.9E1 & 0.13 & 42  \\
        & 2.0 & 6.3E2 & 4.0E1 & 0.16 & 16  \\
        & 1.0 & 7.8E1 & 2.8E1 & 0.22 &  3  \\
   \hline
  \end{tabular}
 \end{center}
\end{table}

As seen from Table~\ref{time scale table} and Fig.~\ref{time scale figure},
since the time scales vary considerably as a function of the separation,
we can divide the evolution of binary star systems due to gravitational 
wave emission into three distinct stages: 
1) the quasi-equilibrium stage where $\tau_{\rm GW} \gg \tau_{\rm rot}$ 
or $d \go R_*$,
2) the merging stage where $\tau_{\rm GW} \sim \tau_{\rm rot}$ 
or $d \sim R_*$,
and 
3) a neutron star or a black hole formation stage where at first 
$\tau_{\rm GW} \sim \tau_{\rm rot}$ or $d \sim R_*$ and 
later $\tau_{\rm GW} \go \tau_{\rm rot}$ or $d \lo R_*$.
In stage 1, gravitational wave emission hardly affects the orbital motion 
during several rotational periods, while in stages 2 and 3 an appreciable 
change of the orbital motion and/or configuration occurs in one or less 
orbital period or on a dynamical time scale.  

Therefore, hydrodynamical simulations are most appropriate for investigations 
of stages 2 and 3 but not for stage 1.  For stage 1, it is proper to assume 
that binary stars are in quasi-equilibrium states because there is little 
change of the configuration and the orbital motion of the binary star 
system.  Moreover, we need to take the structure of the stars into account
for the state with rather small separations $d/R_* \lo 2$ 
where this quasi-equilibrium approach may be still applied.  
Evolution of a binary star system in this 
stage is well approximated by an equilibrium sequence of a binary with a
{\it finite size}, if we take account of proper constraints in  
construction of the sequence.

\subsection{Evolution of velocity fields and equilibrium sequences}
One equilibrium state of a binary star system can be specified by
the following quantities: 1) masses of the stars, 2) the value of the 
separation, 3) the equation of state and 4) rotation laws
of the stars including the internal motions.

Concerning the first two, we can freely choose those values. Although the 
equation of state for the realistic neutron star matter has not been
fully understood for the high density region above the nuclear density,
we will be able to try several sets of equations of state.

However, it is not easy to specify the rotation law for binary star systems
because the rotation law depends on the initial state, the physical state 
of the matter, the evolutionary time and so on. In this paper, we will
consider two extreme cases for viscosity: 1) strong viscosity and 2)
weak viscosity.  If the effect of viscosity is strong enough, the internal
flow would synchronize with the orbital rotation. On the other
hand, if it is weak, the initial state of the flow would be maintained.

As Kochanek\cite{ko92} and Bildsten and Cutler\cite{bc92} have shown,
the situation for neutron stars becomes very simple. 
The effect of viscosity for realistic neutron stars is so small that
it may not affect the rotation law of binary neutron star systems.
This implies that the vorticity of the neutron star system is conserved 
during evolution because the radiation reaction force due to gravitational
wave emission is the potential force at the leading order.  

Furthermore, since values of the angular velocity and the vorticity, 
$\zeta$, of neutron stars will increase as the orbit shrinks, the following
relations will be established at the final state of the quasi-equilibrium 
stage:
\begin{eqnarray}
|\Omega_f| & \gg & |\Omega_i|,  \quad |\zeta_i| \ \ , \\
|\zeta_f|  & \gg & |\Omega_i|,  \quad |\zeta_i| \ \ , 
\end{eqnarray}
where quantities with suffix $_i$ or $_f$ denote those at the initial
state or at the final state, respectively.  Here $\zeta_i$ and $\zeta_f$
are the vorticities in the rotating frame.  Since the vorticity in the 
inertial frame $\zeta_0$ is conserved, we can obtain the following
relation:
\begin{equation}
|\zeta_0| \ll |\Omega_f|, \quad |\zeta_f| \ \ . 
\end{equation}
This means that the vorticity in the inertial frame can be considered
to be negligibly small compared with the angular velocity and the vorticity 
in the rotating frame at the final state of the quasi-equilibrium
evolution stage. In other words, we may consider that flow fields
of binary neutron star systems can be regarded to be {\it irrotational}.

Therefore, in this paper, we will treat two extreme cases: 1) synchronously
rotating binary star systems and 2) irrotational binary star systems.

In principle, we can obtain equilibrium sequences by connecting equilibrium
states with the specified rotation law. However, in order to follow 
quasi-stationary evolution of binary star systems due to gravitational wave 
emission by making use of a quasi-equilibrium sequence, we need to consider 
conserved quantities during evolution.  For binary neutron star systems, one 
conserved quantity is of course the vorticity in the inertial frame as 
mentioned above. For binary star systems emitting gravitational waves, 
the baryon number or the baryon mass is the second conserved quantity because 
matter does neither flow out from nor flow into the system.  Therefore, we 
can construct equilibrium sequences by connecting equilibrium states with 
the same baryon mass and use them as quasi-stationary evolution sequences 
due to gravitational wave emission.

\subsection{Characteristic feature of equilibrium sequences}
Equilibrium sequences can be characterized by several sets of physical 
quantities such as the total angular momentum -- separation relation,
the angular velocity -- separation relation, the energy -- separation 
relation and so on.  In this paper, we will choose the total angular
momentum -- separation relation as the typical relation which
reflects the essential feature of the equilibrium sequence or,
in other words, of the quasi-stationary evolution sequence of the binary
star system due to gravitational wave emission.

In order to show the characteristic feature of the equilibrium
sequence, we will assume that masses of two stars are the same and 
that the stars rotate in a circular orbit. The total angular momentum 
$J_{\rm tot}$ is divided into two parts as follows:
\begin{equation}
J_{\rm tot} = J_{\rm orb} + J_{\rm spin} \ \ ,
\end{equation}
where $J_{\rm orb}$ and $J_{\rm spin}$ are the orbital angular momentum
and the spin angular momentum, respectively.  If we assume that the stars
are congruent, they are expressed as:
\begin{eqnarray}
J_{\rm orb}  & = & {1 \over 4} (2M)^{3/2} A^{1/2} \ \ , \\
J_{\rm spin} & = & 2 (2 M)^{1/2} I A^{-3/2} 
        + 2 \int \rho r \sin \theta v_{\varphi} d^3 r \ \ , \\
J_{\rm tot}  & = & {1 \over 4} (2M)^{3/2} A^{1/2} 
        + 2 (2 M)^{1/2} I A^{-3/2} 
        + 2 \int \rho r \sin \theta v_{\varphi} d^3 r \ \ , 
\end{eqnarray}
where we have used the relation
\begin{equation}
\Omega^2 A^3 = 2 M \ \ . 
\end{equation}
Here $I$, $\rho$ and $v_{\varphi}$ are the moment of inertia, the 
density and the $\varphi$-component of the internal velocity of the 
component star in the rotating frame, respectively, and $(r, \theta, \varphi)$
are the coordinates whose origin coincides with the center of mass of the 
component star.

From these expressions, we can see the dependency of the angular
momentum on the separation as well as on the internal structure of the 
star.  First, as the separation decreases, the orbital angular
momentum decreases as $A^{1/2}$, while the spin angular momentum
increases as $A^{-3/2}$. Second, the internal structure of the 
finite-size star is reflected in the second and third terms of 
$J_{\rm tot}$ through $I$ and $v_{\varphi}$. The moment of inertia 
depends on the matter distribution which is governed by its self-gravity,
the tidal force from the companion star, the orbital motion, the internal 
flow including rotation and the equation of state. The internal velocity 
$v_{\varphi}$ is determined from the initial state and evolution of the 
velocity field.

\begin{figure}
  \epsfysize = 5cm
  \centerline{\epsfbox{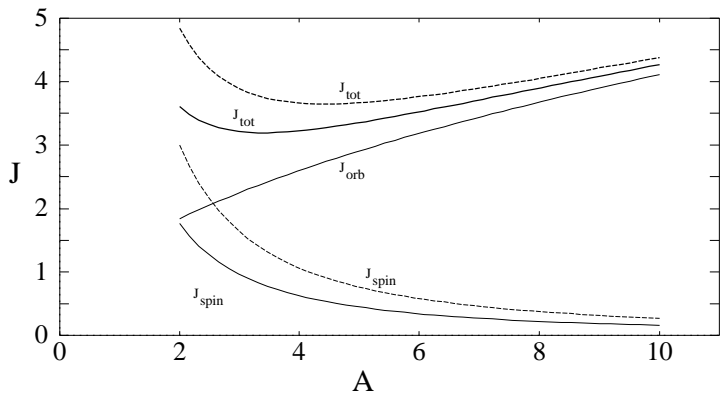}}
  \caption{Schematic figure of the angular momentum $J$ --
           separation $A$ relation. The effect of the moment of inertia
           is shown. Thin solid curves correspond to $J_{\rm orb}$ and 
           $J_{\rm spin}$ and thick solid curve to $J_{\rm tot}$ for a 
           certain value of $I$. Dashed curves correspond to those for 
           larger $I$.}
  \label{I-effect}
\end{figure}

\begin{figure}
  \epsfysize = 5cm
  \centerline{\epsfbox{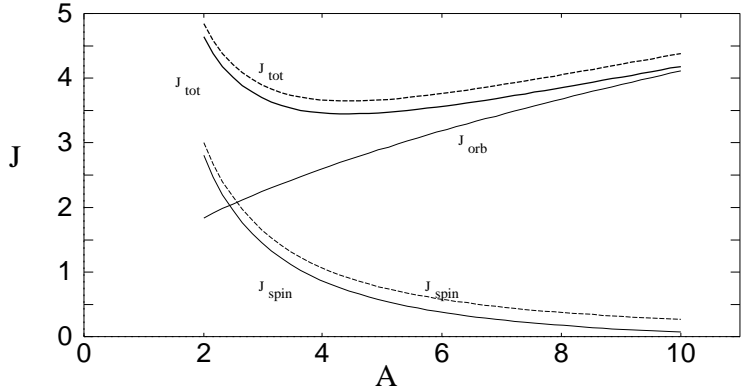}}
  \caption{Same as Fig.~\protect\ref{I-effect} but for the effect of
           the internal velocity. Dashed curves correspond to the
           sequence without internal velocity fields. Solid curves
           are those for models with internal velocity $v_{\varphi} < 0$.}
  \label{v-effect}
\end{figure}

The first feature implies that there can exist a minimum
angular momentum state for a certain value of the separation,
say $A_{\rm min}$. From the second feature the effect of the internal 
structure changes the spin angular momentum but keeps the orbital 
angular momentum unchanged. In particular, the change of the moment of inertia
will result in the shift of the value of $A_{\rm min}$ because the
moment of inertia affects through the term which behaves as $A^{-3/2}$.
Thus, if $I$ increases (decreases), $A_{\rm min}$ becomes larger (smaller). 
The effect of the internal flow is not so drastic. If we consider
the velocity field which flows in the counter direction with respect to the 
orbital motion, i.e. $v_{\varphi} < 0$, the spin angular momentum is 
decreased almost evenly for all separations so that the total angular
momentum decreases but the position of $A_{\rm min}$ is
hardly shifted.  These features are shown in Figs.~\ref{I-effect}
and \ref{v-effect}.


\section{Critical states along the equilibrium sequence of
binary star systems}
When we consider evolutionary sequences of binary star systems, it is 
important to know whether each model along the sequences is stable or not 
against a certain kind of perturbations. If some kind of instability sets in 
at some point, equilibrium configurations beyond that point do not exist in 
realistic situations. Even if instability does not occur, at some stage there 
may occur some kind of significant change of configurations.  In this section 
we will discuss such kind of critical states which will appear along 
evolutionary sequences.

\subsection{Contact state for congruent stars}
For congruent stars, if the separation decreases, it is most probable that 
two binary stars come to contact by considering the symmetry of configurations
in equilibrium states.  In order to find this critical state, we need to treat 
deformation of configurations very accurately.  We will use $A_c$ or $d_c$
for the separation of this contact phase or the half of the separation, 
respectively. After this contact phase, its configuration is no more a binary 
state but a single body is formed.  

\subsection{Roche lobe filling state (cusp formation state)}
If masses of two stars are different or if configurations of two stars
are different such as a black hole -- neutron star system or if there exist
internal motions, decrease of the separation results in a Roche lobe filling 
state. Here we use the word ``Roche lobe'' in its generalized sense because 
models can include internal flows.  Since there is a cusp point
on the surface of the Roche lobe, a cusp shape appears on
the stellar surface, too.  This state is denoted by the separation
$A_{\rm OF}$ or the half of the separation $d_{\rm OF}$. 
Here the subscript $_{\rm OF}$ denotes that the mass begins to overflow
from the Roche lobe filling star to the other star.

\subsection{Roche limit state (minimum separation state)}
The Roche limit is defined as the minimum separation for a binary system.
Within the Roche limit there are no equilibrium states for the component 
stars. For a spherical star, the Roche limit state coincides with the state 
at the maximum angular velocity. However, if we take deformation of stars 
into account, this is no more the case. For some binary configurations, 
equilibrium states with larger angular velocity than that of the Roche limit 
state can exist.  The Roche limit and its half distance are denoted as
$A_{\rm RL}$ and $d_{\rm RL}$, respectively.

\subsection{Turning point of equilibrium sequence}
As discussed before, along the equilibrium sequence of binary star systems,
there may appear a state where the total angular momentum becomes minimum.
Since gravitational waves carry away the angular momentum from the system,
the minimum state of the angular momentum denotes a critical state for
equilibrium sequences. 
For synchronously rotating binary star sequences, 
this state corresponds to the onset of {\it secular instability} due to 
viscosity. On the other hand, for irrotational binary star sequences,
{\it dynamical instability} of the orbital motion sets in at this point.
We will use $A_{\rm dyn}$ and $d_{\rm dyn}$ for the separation and the 
half of it at this state, respectively.
 
\subsection{Dynamical instability against collapse}
For compact stars such as white dwarfs or neutron stars, we need to
consider the stability of a single star to collapse. This instability
sets in at the state of the maximum gravitational mass along the equilibrium
sequence with the same distribution of the specific angular momentum
for barotropes in Newtonian gravity.~\cite{bb74} \ 
(In general relativity situation is somewhat different.\cite{fis88}) \  
The criterion of this instability can be written as follows:
\begin{equation}
\rho_{\rm max} > \rho_{\rm max,c} \ \ ,
\end{equation}
where $\rho_{\rm max}$ and $\rho_{\rm max,c}$ are the maximum density
of the star and the density for the maximum gravitational mass state,
respectively. It is not easy to find this critical configuration 
exactly because the equilibrium sequences for compact binary star systems 
are not those with the distribution of the specific angular momentum kept 
constant.  However, roughly speaking, if the maximum density of the star 
increases as the separation decreases, there arises a possibility of onset 
of this instability for each component star.

\subsection{General relativistic instability}
In this paper, although we treat equilibrium configurations only in 
Newtonian gravity, we should point out that there is a critical state
due to general relativistic instability of the orbital motion, i.e. the 
innermost stable circular orbit.  The critical separation or the
distance for this instability is denoted as $A_{\rm GR}$ or $d_{\rm GR}$.
Here 
\begin{equation}
d_{\rm GR} = 6 M_{\rm tot} + \Delta \ \ ,
\end{equation}
where $M_{\rm tot}$ is the total mass and $\Delta$ is a positive quantity 
which depends on the general relativistic structure of the system.

\subsection{Classification of binary equilibrium sequences by  
critical states}
By considering the order of appearance of these critical points along
the equilibrium sequence, we can classify the quasi-stationary evolution of
binary star systems into three types with several subtypes from the 
topology of the equilibrium sequence on the total angular momentum -- 
separation plane.

\vskip 5truemm

\begin{description}
\item[Type 1a:] \ \  The equilibrium sequence terminates at either 
                     $A_c$ or $A_{\rm OF}$. There is no turning point.
\item[Type 1b:] \ \ The equilibrium sequence terminates at $A_{\rm RL}$.
                    There is no turning point. 
\item[Type 2:] \ \ The equilibrium sequence terminates at either $A_c$, 
                   $A_{\rm OF}$ or $A_{\rm RL}$ but the first critical point 
                   along the equilibrium sequence is the turning point.  
                   Thus the following condition is satisfied:
                   $ A_{\rm dyn} > A_c, A_{\rm OF}, A_{\rm RL}$.
\item[Type 3:] \ \ The equilibrium sequence terminates at either $A_c$ or 
                   $A_{\rm OF}$.  The first critical point along the 
                   equilibrium sequence is the turning point.  Thus the 
                   following condition is satisfied:
                   $ A_{\rm dyn} > A_{\rm RL} $.
\end{description}

\vskip 5truemm

\begin{figure}
  \epsfysize = 5cm
  \centerline{\epsfbox{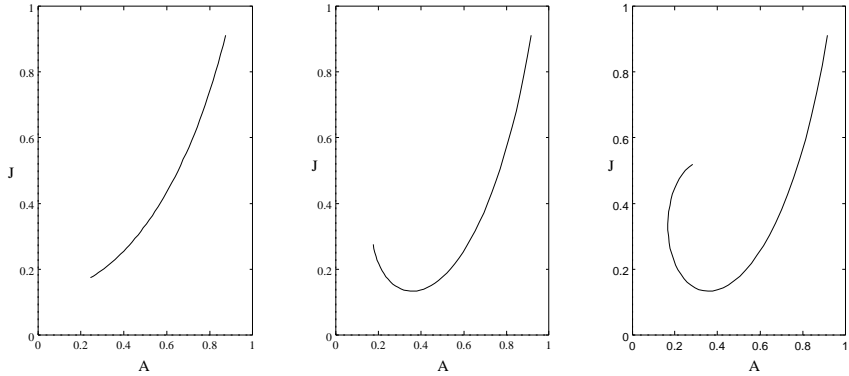}}
  \caption{The total angular momentum is plotted against the separation
           for equilibrium sequences which belong to Type 1 (left panel),
           Type 2 (middle panel) and Type 3 (right panel). 
           The Type 1 equilibrium sequence terminates at either the contact 
           state, the Roche lobe overflow state (Type 1a) or the Roche 
           limit state (Type 1b).
           The Type 2 equilibrium sequence terminates at the contact 
           state, the Roche lobe overflow state or the Roche limit
           state but there is a turning point along the sequence.
           The Type 3 equilibrium sequence terminates at the contact 
           state or the Roche lobe overflow state. There is a 
           turning point along the sequence. The Roche limit state
           comes between these critical configurations.}
  \label{type 123}
\end{figure}

For quasi-stationary evolution of Type 1a, the final outcome is a
single star or mass overflow from one component star to the other.
For stars of Type 1b, 2 and 3, they suffer from dynamical
disruption eventually. Of course, if the turning points are those for
secular instability, stars will excite internal motion due to viscosity
and can come closer. However, at the end they will finally encounter
the turning point where dynamical instability sets in and suffer from 
dynamical disruption.  Schematic figures of these types are shown in
Fig.~\ref{type 123}. Almost the same analysis has been done by Lai, Rasio
and Shapiro.\cite{lrs934}

\section{Classical results for binary star systems}
Until early of 1980s, it was hard to obtain binary configurations
of compressible gases.  Therefore, structures of binary star systems
had been investigated by assuming that 1) the density is constant,
that 2) the shape of the star is ellipsoid, that 3) the angular velocity
is constant and that 4) the vorticity is also constant.
Under these assumptions three kinds of classical ellipsoids
for binary star systems have been solved (see e.g. Ref.~\citen{ch69}).

\subsection{Roche ellipsoid: synchronously rotating sphere 
(or point mass) -- ellipsoid system}
Let us assume that one of the stars is a point mass or a sphere and
that the other star is synchronously rotating with the orbital motion. 
If we include only the leading term of the tidal force from the point
mass or the spherical star, the equilibrium condition for the fluid star
can be expressed as
\begin{equation}
{p \over \rho} = - \phi_1 + {GM_2 \over A^3} 
\left( x^2 - {1 \over2}y^2 - {1 \over 2} z^2 \right) + {1 \over 2} \Omega^2
(x^2+y^2) + {\rm constant} \ \ ,
\end{equation}
where $p$, $\phi_1$, $M_2$ are the pressure, the gravitational potential 
of the fluid star (self-gravity) and the mass of the spherical star, 
respectively. The Cartesian coordinates $(x,y,z)$ are used. If the shape
of the fluid star is ellipsoid, this equilibrium condition can be satisfied
by choosing the angular velocity $\Omega$ appropriately.  This ellipsoid
is called the Roche ellipsoid.

\begin{figure}
    \epsfysize = 5cm
    \centerline{\epsfbox{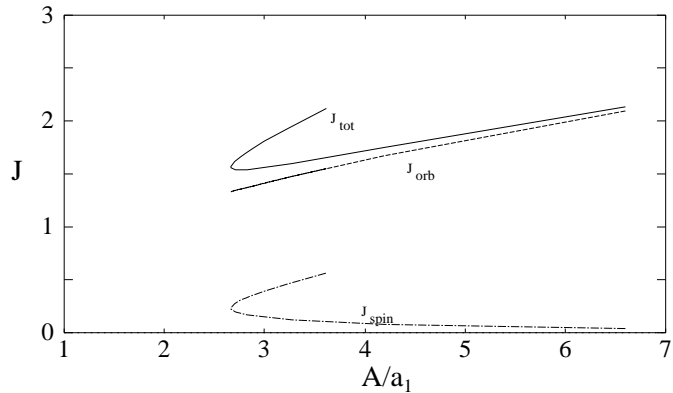}}
    \caption{Angular momentum is plotted against the separation for
             Roche ellipsoids.
             The orbital angular momentum (dashed curve), the spin angular
             momentum (dash-dotted curve) and the total angular momentum
             (solid curve) are shown. The turnaround of the spin curve
             is due to changes of the internal structure.}
    \label{roche}
\end{figure}

For the Roche ellipsoids with $M_1 = M_2$, the angular momentum -- separation
relation is shown in Fig.~\ref{roche}. 
\begin{figure}
    \epsfysize = 5cm
    \centerline{\epsfbox{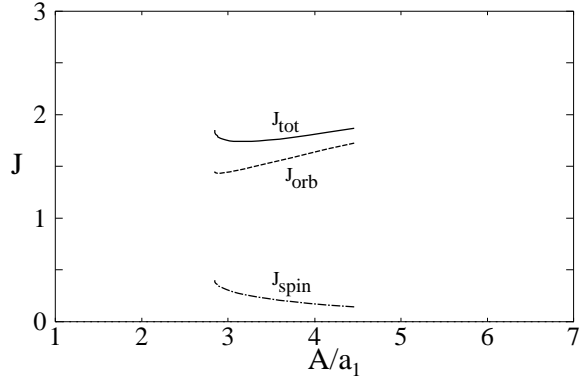}}
    \caption{Same as Fig.~\protect\ref{roche} but for Darwin ellipsoids.
             There is no turnaround of the spin curve.}
    \label{darwin}
\end{figure}
As seen from this figure, the Roche ellipsoidal sequence belongs
to Type 3. The curve of the spin angular momentum shows a turnaround
due to a significant change of the internal structure.  Along this
sequence there appears a turning point where secular instability
sets in.  At this turning point the shape of ellipsoid
is specified by $a_2/a_1 = 0.65$ and $a_3/a_1 = 0.59$ where $(a_1, a_2, 
a_3)$ are three principal axis distances of the ellipsoid.

\subsection{Darwin ellipsoid: synchronously rotating congruent ellipsoids}
If two fluid stars are congruent and synchronously rotating, the equilibrium
condition can be expressed as follows:
\begin{equation}
{p \over \rho} = - \phi_1 + (\beta_1 x^2 + \beta_2 y^2 + \beta_3 z^2) 
+ {1 \over 2} \Omega^2 (x^2+y^2) + {\rm constant} \ \ ,
\end{equation}
where $\beta$'s are constants which can be determined from the leading
term of the tidal force from the secondary star.  For this case,
the ellipsoidal configuration satisfies the above equilibrium condition.
Solutions to this equation are called Darwin ellipsoids. The equilibrium 
sequence is shown in Fig.~\ref{darwin}. The sequence terminates at the 
contact state. Thus the Darwin sequence belongs to Type 3 sequence. 
The Roche limit state is characterized by $a_2/a_1 = 0.62$
and $a_3/a_1 = 0.56$.

\subsection{Roche-Riemann ellipsoid: a sphere (or point mass) --
ellipsoid system with vorticity}
Fluid binary stars with nonzero vorticity were first investigated by 
Aizenman.\cite{ai68}  The system is almost the same as the Roche problem
but in the fluid star there exist internal motions represented by
the vorticity.  The equilibrium condition is written as follows:
\begin{equation}
{p \over \rho} = - \phi_1 - \phi_2 + {1 \over 2} \Omega^2 (x^2+y^2) 
+ {1 \over 2} u^2 + (2 \Omega + \zeta) \Psi + {\rm constant} \ \ ,
\end{equation}
where $\phi_2$, $u$ and $\Psi$ are the gravitational potential from the
spherical star, the velocity of the internal motion in the rotating frame
and the stream function of the internal motion, respectively. This
condition can be satisfied by ellipsoids and ellipsoidal solutions
are called Roche-Riemann ellipsoids.  
According to the results of Aizenman, by introducing the internal
flows, the minimum separation, i.e. Roche limit, can become smaller than
hat for the Roche ellipsoids.  In other words, two stars can come closer
each other.

\section{Recent results for binary star systems in Newtonian gravity}
Although basic properties of binary star systems have been understood by
using classical results mentioned in the previous section, several crucial 
things have remained to be investigated.  First, the constant density
assumption is not valid for realistic binary stars. Second, the 
real shape of deformed stars is not ellipsoidal. Therefore we need to
take these points into account to understand realistic evolution of
compact binary star systems.

\subsection{Ellipsoidal approach}
The effect of the compressibility is introduced by Lai, Rasio and
Shapiro.\cite{lrs934} \  They have investigated polytropic binary
star systems:
\begin{equation}
p = K \rho^{1 + 1/n} \ \ ,
\end{equation}
where $K$ and $n$ are a polytropic constant and the polytropic index,
respectively.  Concerning the shape of binary stars, they have assumed
that it is approximated by ellipsoidal shape. By using the energy
variational principle, they have obtained approximated equilibrium 
configurations.

Their main results are as follows. For simplicity we will call Roche
or Roche-Riemann ellipsoids for the systems consisting of a spherical star 
(or point mass) and a compressible ellipsoid without or with internal 
motions, respectively. Darwin or Darwin-Riemann ellipsoids are used for
the compressible congruent stars without or with internal motions.

For Roche and Roche-Riemann ellipsoids, 
\begin{equation}
A_{\rm dyn} > A_{\rm RL} \ \ .
\end{equation}
This means that for these binary systems secular instability (Roche 
ellipsoids) or dynamical instability (Roche-Riemann ellipsoids) sets 
in before the Roche limit states are reached.

For Darwin ellipsoids, there is a critical value of the polytropic
index as follows:
\begin{equation}
n_{\rm cr}^{\rm D} \approx 2 \ \ ,
\end{equation}
and
\begin{eqnarray}
{\rm for} \ \ n < n_{\rm cr}^{\rm D}  & , \ \ \ & A_{\rm dyn} >  A_{c}    
    \ \ ,  \\
{\rm for} \ \ n \go n_{\rm cr}^{\rm D} & , \ \ \ & A_c {\rm \ \ comes \ \
    first \ \ (no} \ \ A_{\rm dyn}) \ \ .
\end{eqnarray}
This means that for rather stiff equations of state 
($n < n_{\rm cr}^{\rm D}$) 
synchronously rotating compressible binary star systems suffer from
secular instability and develop internal motions. However, for soft
equations of state, stars come to contact without suffering from
any critical phenomena.

For Darwin-Riemann ellipsoids, the situation is as follows.
The critical polytropic index for the Darwin-Riemann ellipsoids is
\begin{equation} 
n_{\rm cr}^{DR} \approx 1.2  \ \ , 
\end{equation}
and
\begin{eqnarray}
{\rm for} \ \ n < n_{\rm cr}^{\rm DR}  & , \ \ \ & A_{\rm dyn} >  A_{c}    
    \ \ ,  \\
{\rm for} \ \ n \go n_{\rm cr}^{\rm DR} & , \ \ \ & A_c {\rm \ \ comes \ \ 
    first \ \ (no} \ \ A_{\rm dyn}) \ \ .
\end{eqnarray}
Thus for rather stiff binary stars ($n < n_{\rm cr}^{\rm DR}$) 
dynamical disruption would occur.

\subsection{Numerical solutions to exact basic equations for binary 
configurations in Newtonian gravity}
Very recently {\it exact equations} for binary configurations in Newtonian 
gravity have been numerically solved by us.\cite{ue989} \ In our treatment, 
we assume 1) the polytropic equation of state, 2) the stationary state in the 
rotating frame, 3) inviscid fluid and 4) irrotational flows in the inertial 
frame. We do not assume the shape of configurations but we solve for the 
shape of configurations together with other quantities.

Basic equations are written as follows:
\begin{equation}
\int {d p \over \rho} + \phi + {1 \over 2} | \nabla \Psi |^2
 - (\mib{\Omega} \times {\mib r}) \nabla \Psi = constant \ \ ,
\end{equation}
\begin{equation}
  \phi = - \int {\rho({\mib r}^{'}) \over |{\mib r} - {\mib r}^{'}|} 
      d^3{\mib r}^{'} \ \ ,
\end{equation}
\begin{equation}
  \Delta \Psi = n(\mib{\Omega} \times {\mib r}- \nabla \Psi) 
  \cdot {\nabla \Theta \over \Theta} \ \ ,
\end{equation}
where $\Theta$ is defined as
\begin{equation}
p = K \Theta^{1 + n} \ \ .
\end{equation}

Boundary conditions for these equations are as follows:
\begin{eqnarray}
\Theta(r_s(\theta, \varphi)) & = & 0 \ \ {\rm for \ \ the \ \ matter} \ \ , \\
{\mib n}_s \cdot {\mib v}_s  & = & 0 \ \ {\rm for \ \ the \ \ flow} \ \ ,
\end{eqnarray}
where $r = r_s(\theta, \varphi)$, ${\mib n}_s$ and ${\mib v}_s$ are
the surface of the star, the normal vector to the surface and the velocity
vector along the surface, respectively.

\begin{figure}
    \epsfysize = 5cm
    \centerline{\epsfbox{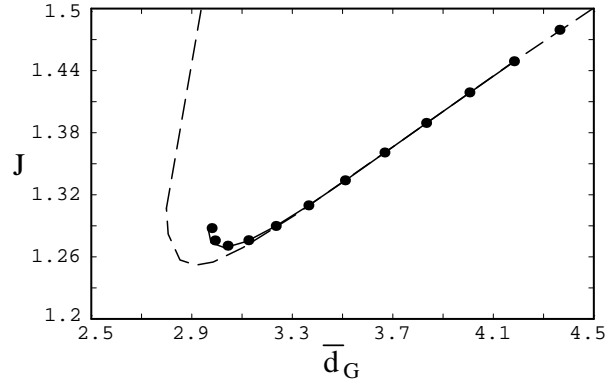}}
  \caption{The total angular momentum -- separation relation for
           the irrotational Roche-Riemann incompressible sequence. 
           The result from the ellipsoidal approximation\protect\cite{lrs934} 
           (dashed curve) and that of numerical 
           computations\protect\cite{ue989} (solid curve and dots) are 
           shown. The solid curve and the dots are computed by 
           independent numerical methods. }
  \label{exact roche-riemann}
\end{figure}

\begin{figure}
    \epsfysize = 5cm
    \centerline{\epsfbox{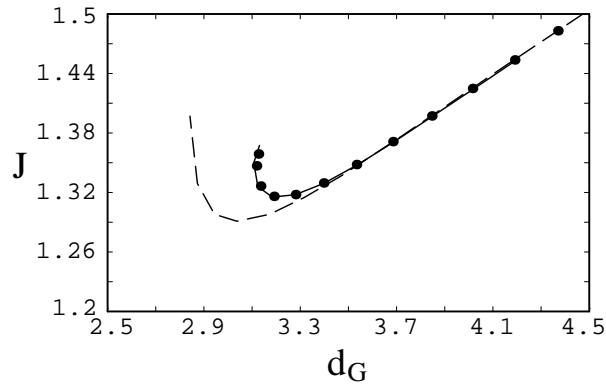}}
  \caption{Same as Fig.\protect\ref{exact roche-riemann} but for
           the irrotational Darwin-Riemann incompressible sequence.}
  \label{exact darwin-riemann}
\end{figure}

In order to show only the effect of the shape and the structure,
we will present the results for $n = 0$ polytropes or incompressible fluids.
In Figs.\ref{exact roche-riemann}-\ref{exact darwin-riemann}, results 
from the ellipsoidal approximation\cite{lrs934} and those of numerical 
computations\cite{ue989} are shown. As seen from these figures, the exact
treatment gives larger values of the moment of inertia and so the separation
at the turning point shifts to larger value compared with that of the
ellipsoidal approximation. 

\begin{figure}
  \epsfxsize = 8cm
  \centerline{\epsfbox{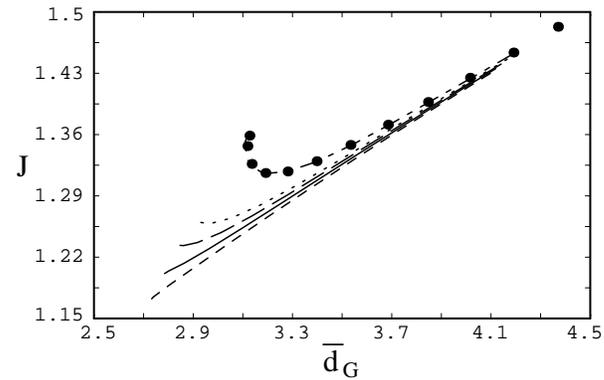}}
  \caption{The total angular momentum -- separation relations for several
           polytropes of irrotational Darwin-Riemann sequences. 
           Polytropes with $n=0.0$ (dash-dotted), $n=0.5$ (dotted),
           $n=0.7$ (long dashed), $n=1.0$ (solid) and $n=1.5$ (short dashed)
           are shown.}
  \label{sequences for eDR}
\end{figure}

For numerical solutions of irrotational Darwin-Riemann compressible 
configurations, there is a critical polytropic index $n_{\rm cr}^{\rm eDR}$
as seen in Fig.\ref{sequences for eDR} (eDR means ``exact'' Darwin-Riemann). 
Therefore, irrotational polytropic binary star systems
can be classified as follows:
\begin{equation} 
n_{\rm cr}^{eDR} \approx 0.7  \ \ , 
\end{equation}
and
\begin{eqnarray}
{\rm for} \ \ n < n_{\rm cr}^{\rm eDR}  & , \ \ \ & 
    A_{\rm dyn} >  A_{\rm OF}  \ \ ,  \\
{\rm for} \ \ n \go n_{\rm cr}^{\rm eDR} & , \ \ \ & A_{\rm OF} 
    {\rm \ \ comes \ \ first \ \ (no} \ \ A_{\rm dyn}) \ \ .
\end{eqnarray}
As discussed before, the critical value $n_{\rm cr}^{\rm eDR} \approx 0.7$
is much smaller than the critical value $n_{\rm cr}^{\rm DR} \approx 1.2$  
obtained from the ellipsoidal approximation. This is an important 
difference because the effective polytropic index for realistic neutron
stars is estimated $0.5 \sim 1.0$. Thus realistic neutron stars may not 
suffer from dynamical instability during evolution due to gravitational wave 
emission but result in mass overflow.

\begin{figure}
  \epsfysize = 5cm
  \centerline{\epsfbox{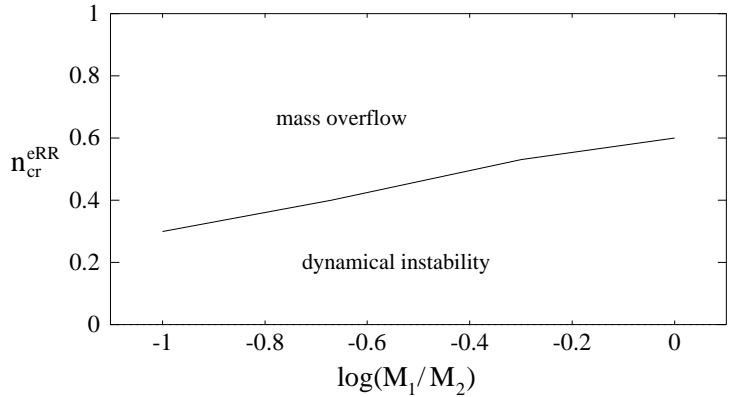}}
  \caption{The critical polytropic index is plotted against the mass
           ratio $\log (M_1/M_2)$ for irrotational Roche-Riemann compressible
           configurations. If the polytropic index is smaller than the
           critical polytropic index, the polytropic star would reach
           dynamically unstable state and disruption would occur. If the
           polytropic index is larger than the critical value, the mass
           overflow to the other star would begin.}
  \label{sequences for eRR}
\end{figure}

\begin{figure}
    \epsfysize = 5cm
    \centerline{\epsfbox{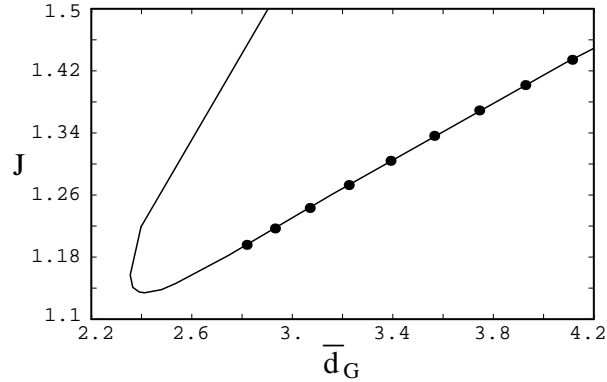}}
  \caption{Same as Fig.\protect\ref{exact roche-riemann} but for
           the irrotational Roche-Riemann $n=1$ polytropic sequence. 
           The results from the ellipsoidal approximation\protect\cite{lrs934} 
           (solid curve) and that of numerical 
           computations\protect\cite{ue989} (dots) are shown. }
  \label{n=1 roche-riemann}
\end{figure}

As for the numerical solutions of irrotational Roche-Riemann compressible 
configurations, critical polytropic indices $n_{\rm cr}^{\rm eRR}$
depend on the mass ratio of the stars.  When 
$ n > n_{\rm cr}^{\rm eRR}$, there appear no turning points and so
binary star systems would result in mass overflow.
If $ n \lo n_{\rm cr}^{\rm eRR}$, turning points appear and
dynamical disruption would occur.  In Fig.\ref{sequences for eRR}, 
the critical polytropic index is shown for different mass ratios for 
irrotational Roche-Riemann compressible binary star systems.
This result is clear contrast to that of the ellipsoidal approximation 
because solutions from the ellipsoidal approximation always give a turning 
point along the sequences as shown in Fig.\ref{n=1 roche-riemann}.

\section{Discussion and Summary}
\subsection{Discussion}
In this paper we have focused on the equilibrium sequences of binary star 
systems in Newtonian gravity. However, for compact binary star systems such 
as neutron star -- neutron star binary systems or neutron star -- black hole 
binary systems, the role of general relativity is important.  The internal 
structure and the orbital nature are affected by general relativity.  
Concerning the internal structure, general relativity tends to make the matter 
softer or increase the ``effective'' polytropic index so that the mass will 
concentrate towards the central region. This implies that mass overflow will 
be more likely to occur. 

As for the orbital motion, from the perturbative analysis of the black 
hole -- neutron star binary system, the following relation for the
innermost stable circular orbit has been obtained:\cite{kww92,lrs944}
\begin{equation}
d_{\rm GR} = 6 M_{\rm tot} + 4 \mu \ \ ,
\end{equation}
where $\mu$ is the reduced mass:
\begin{equation}
\mu \equiv = { M_1 M_2 \over M_1 + M_2 } \ \ .
\end{equation}
From this radius, we can roughly estimate that
\begin{equation}
d_{\rm OF} ( \approx d_{\rm RL} ) > d_{\rm GR} \ , \ \ \
{\rm for} \ \ {M_1 \over M_2} \go 0.25, \  0.5 \lo n \lo 1.0 
\ {\rm and } \  {M_1 \over R_*} \lo 0.15 \ \ .  
\end{equation}
This means that the mass overflow radius is reached before the general 
relativistic effect begins to work.

Concerning the critical radius of the dynamical instability, 
Shibata\cite{sh97} has shown by using the post-Newtonian approximation that
\begin{equation}
  d_{\rm dyn} \ ({\rm for \ post-Newtonian \ case}) 
< d_{\rm dyn} \ ({\rm for \ Newtonian \ case}) 
\ \ .
\end{equation}
Thus the qualitative results about the critical radii obtained from the 
Newtonian analysis would not be changed even in general relativistic analysis,
although the values are necessarily different.

We need to comment on the behavior of the maximum density of the individual
star along the quasi-stationary evolutionary sequence. In general relativistic
computations, Wilson, Mathews and Maronetti\cite{wmm96} have shown that
the maximum density increases as the system evolves. However, recently,
Bonazzola, Gourgoulhon and Marck\cite{bgm99} and Uryu and Eriguchi\cite{ue99}
have shown that the central density of the component star of the irrotational 
binary star systems in general relativity decreases as the binary star system
evolves, although at large separations there is some stage where the central 
density increases slightly.

\begin{figure}
  \epsfysize = 7cm
  \centerline{\epsfbox{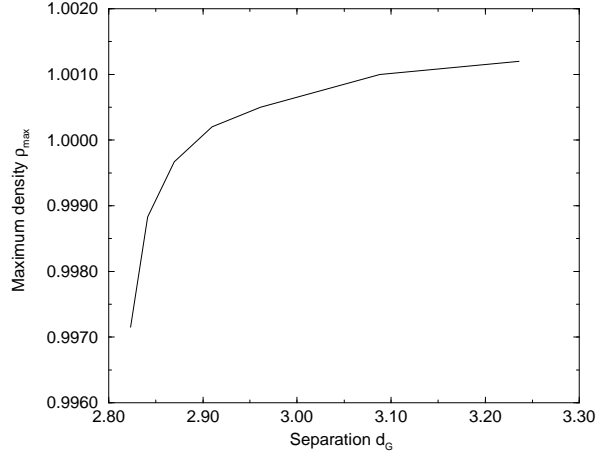}}
  \caption{The maximum density is plotted against the separation
           for $n = 1$ irrotational Darwin-Riemann configuration systems.
           The maximum density is normalized by the maximum density 
           of the spherical star.}
  \label{maximum density}
\end{figure}

Our Newtonian computations also show that the maximum density decreases as 
evolution proceeds. This is shown in Fig.\ref{maximum density}.
For Newtonian polytropes, Taniguchi and Nakamura\cite{tn99} have analyzed 
irrotational binary star systems of $n = 1$ polytropes perturbationally and 
found that
\begin{equation}
 \left( {\rho_{\rm max,i} - \rho_{\rm max} \over \rho_{\rm max,i} } \right)
= \frac{45}{2\,(M_1/M_2)^2\,\xi_1^2} \left(\frac{a_0}{A}\right)^6 \ \ ,
\end{equation}
to the order of ${\cal O}(1/A^6)$, 
where $\rho_{\rm max,i}$ and $a_0$ are the maximum density and the radius 
for the initial spherical star, respectively.  Here $\xi_1$ equals to $\pi$ 
for $n=1$ case.  
Therefore, it is very likely that the maximum mass will not increase
significantly. In other words, the component star would not suffer from
dynamical instability to collapse.

\subsection{Summary}
We have reviewed properties of classical and/or approximated equilibrium 
sequences of Darwin, Darwin-Riemann, Roche and Roche-Riemann problems in 
Newtonian gravity in the context of the final evolution of compact binary 
star systems due to gravitational wave emission.  

We have presented the results of full numerical computations of equilibrium 
sequences of irrotational binary star systems in Newtonian gravity.  From the 
analysis of critical points along these equilibrium sequences, we have shown 
the possibility that evolution of compact binary neutron star systems would 
result in mass overflow instead of dynamical disruption for a certain 
parameter ranges.

\section*{Acknowledgements}
A part of the numerical computations has been carried out at the 
Astronomical Data Analysis Center of the National Astronomical 
Observatory, Japan.  One of us (KU) would like to thank Prof. J.C. 
Miller for discussions and continuous encouragements. He would also 
like to thank Prof. D.W. Sciama and A. Lanza for his warm hospitality 
at SISSA and ICTP.

%
%

%

\end{document}